\documentclass[structabstract]{aa}

\usepackage{epsfig}
\usepackage{graphicx}
\usepackage[latin1]{inputenc}
\usepackage{txfonts}
\usepackage{rotating}
\usepackage{natbib}

\begin{document}

   \title{Matching microlensing events with X-ray sources}

   \subtitle{}

   \author{N. Sartore
          \inst{1,2},
          A. Treves
          \inst{2,3},
          }

   \institute{INAF - Istituto di Fisica Spaziale e Fisica Cosmica,
              via Bassini 15, 20100, Milan, Italy
              \email{sartore@iasf-milano.inaf.it}
         \and
             Dipartimento di Fisica e Matematica, Università dell'Insubria,
             via Valleggio 11, 22100, Como, Italy
         \and
             Affiliated to INAF and INFN
             }

   \date{Received ...; accepted ...}

 \abstract
    {}
    {The detection of old neutron stars and stellar mass black holes in isolation is one of the cornerstones of compact
object astrophysics. Microlensing surveys may help on this purpose since the lensing mechanism is independent of the emission
properties of the lens. Indeed, several black hole candidates deriving through microlensing observations have been reported 
in the literature. The identification of counterparts, especially in the X-rays, would be a strong argument in favor 
of the compact nature of these lenses.}
    {We perform a cross-correlation between the catalogs of microlensing events by the OGLE, MACHO and MOA teams, 
and those of X-rays sources from XMM-Newton and Chandra satellites. Based on our previous work, we select only microlensing 
events with duration longer than one hundred days, which should contain a large fraction of lenses as compact objects. 
Our matching criterion takes into account the positional coincidence in the sky.}
    {We find a single match between a microlensing event, OGLE 2004-BLG-81 ($t_E \sim 103$ days), and the X-ray 
source 2XMM J180540.5-273427. The angular separation is $\sim 0.5$ arc-seconds, i.e. well inside the $90\%$ error box of the 
X-ray source.
The hardness ratios reported in the 2XMM catalog would suggest a hard spectrum with a peak between 2 and 4.5 keV 
or a softer but highly absorbed source. Moreover the microlensing event is not fully constrained, and possible association 
of the source star with a flaring cataclysmic variable or a RS Canum Venaticorum-like star have been 
suggested as well.}
    {The very small angular separation (within uncertainties) is a strong indicator that 2XMM J180540.5-273427 is the X-ray
counterpart of the OGLE event. However, the uncertainties on both the nature of lensed system and the lens itself
challenge the interpretation of 2XMM J180540.5-273427 as the first confirmed isolated black hole identified so far. 
In any case, it probes the diagnostic capacity of microlensing surveys and open the path for further identifications 
of black hole or neutron star candidates.}


\maketitle

\section{INTRODUCTION}

The Galactic population of neutron stars and and black holes (NSs and BHs hereafter) is expected to count as much as 
$10^8 - 10^9$ objects \citep[e.g.][]{kk2008}. This population is almost undetected, with the exception of few thousands of 
sources, mainly isolated young NSs emitting as radio/X-ray/gamma-ray pulsars and NSs and BHs in X-ray binaries.

Observing the large majority of old, likely isolated objects is difficult since the proposed recycling mechanism, 
i.e. residual accretion from the interstellar medium \citep[e.g.][]{ors1970, tc1991}, is hampered by many factors, 
like the large velocities \citep[e.g.][]{treves2000, popov2000, sartore2010}, the presence of the magnetic field in NSs 
\citep[e.g.][]{toropina2003, perna2003} and, for BHs, the small radiation efficiencies of the accretion process 
\citep[e.g.][]{ak2002, bk2005}. Nevertheless, isolated old remnants (NSs and BHs) can certainly act as 
gravitational lenses.

Gravitational microlensing (ML) events occur when light from a star is deflected and amplified by a foreground mass 
\citep{paczynski1986}. Indeed, several thousands of ML events have been observed so far \citep[see][for a review]{moniez2010}. 
Since most of the events are due to stars with mass less than one solar mass, NSs and BHs contribute essentially 
to the long duration events, the longest ones being associated to BHs \citep[e.g.][]{mao2002, bennett2002}. 
According to \citeauthor{bennett2002}, the number of long duration events is in excess of that inferred from theoretical models 
\citep[e.g.][]{gould2000, wm2005}.
Recently \cite{st2010} showed that this excess can be explained assuming that BHs receive large natal kicks at birth, 
of the same magnitude of those observed in NSs \citep[e.g.][]{hobbs2005, gualandris2005}. 
Under this assumption, $\sim 40$ percent of the events longer than 100 days are due to BHs.

Thus, microlensing surveys may have already detected tens of events related to NSs and BHs. However, the nature of candidate 
BH lenses has to be assessed with independent methods. The mass estimated from ML events is in general affected 
by large uncertainties, given the degeneracy of the time-scale with the input parameters like the mass of the lens 
and the geometry of the source-lens-observer system, i.e. the relative positions and velocities.
On the other hand, the detection of X-rays coming from the position of the ML event would be a strong hint 
for the presence of a compact object.
A search for the X-ray counterpart of one of the most promising BH candidates detected through ML has been performed by 
\citet[][see also \citealt{ak2002}]{nucita2006}. However, a 100 ks pointing with XMM-Newton gave no result down to a 
limiting flux $\sim 10^{-15}\, \rm erg\,s^{-1}$ in the $0.2-10$ keV energy range.

In this work we present a systematic cross-correlation analysis of long duration ML events with the X-ray catalogs 
of XMM-Newton and Chandra satellites, which appeared recently. 
The paper is organized as follows: Sect.\ref{spectra} shortly describes the expected spectra of NSs and BHs 
accreting at low rates. In Sect \ref{method} we report on the list of ML events and catalogs of X-ray sources used 
to find possible counterparts. A description of the cross-correlation method is also given. 
In Sect 4 we present our results and we briefly discuss on a source resulting form the cross-correlation procedure. 
Finally in Section \ref{discussion} we summarize the main points of the paper and discuss the possibilities 
of finding more remnant candidates with ML.

\section{EXPECTED X-RAY SPECTRA OF ACCRETING REMNANTS}\label{spectra}

It is commonly accepted that isolated NSs and BHs accreting from the ISM should radiate mostly in the X-ray band. 
For instance, the spectrum emitted by a NS accreting at low rate could be crudely approximated to a Planckian, 
with a temperature of few hundreds elettronvolt \citep{ors1970, tc1991, bm1993}.
More detailed calculations \citep{zampieri1995, zane2000} showed that the spectrum emerging from an accreting NS, 
either magnetized or non-magnetized, could have substantial deviations from a blackbody and in particular 
the resulting spectrum would be harder than the blackbody emission at the same accretion rate. 
However, the overall behavior would still resemble a Planckian-like spectrum with peak emission in the soft X-rays.

The situation for BHs is quite different since, at variance with NSs, the efficiency of conversion of accreted matter 
into radiation is expected to be lower. 
Models of ISM accretion onto isolated BHs have been developed by \cite{ak2002, bk2005, mapelli2006}. 
The underlying physics is assumed to be similar to that of BH candidates in X-ray binaries but the exact details 
are poorly constrained. Nevertheless, the emerging spectra should have to have both thermal-like and non thermal components, 
as observed in X-ray binaries.

The X-ray luminosities of accreting NSs/BHs would depend heavily on the relative velocity of the compact object (NS/BH) 
with respect to the surrounding medium \citep[e.g.][]{bondi1952}. 

\begin{equation}\label{eqn-lx}
L_X \sim 10^{31}\, \eta\, \Big(\frac{n}{1\, \rm cm^{-3}}\Big)\,
\Big( \frac{v}{10\, \rm km\,s^{-1}} \Big)^{-3}\, \Big( \frac{M}{1\, M_\odot} \Big)^{3}\, \rm erg\,s^{-1}\,,
\end{equation}\bigskip

\noindent where $\eta$ is the efficiency of the X-ray emission, $n$ is the number density of the ISM, 
$v$ and $M$ are the relative velocity with respect to the surrounding medium and the mass of the compact objects, 
respectively. 
Old, but still fast moving NSs are expected to populate the Galactic disk with velocities of $\sim 200\, \rm km\,s^{-1}$ 
\citep[e.g.][]{sartore2010}. The resulting luminosity is of the order of $\sim 10^{27} - 10^{28}\, \rm erg\,s^{-1}$. 
For BHs, the effect of their higher mass is counterbalanced by the lower efficiency and thus the nominal X-ray luminosities 
are approximately of the same order of accreting NSs, see however the paper of \citeauthor{mapelli2006}.
Also, when comparing with observations one has to take into account the absorption effect, which are severe in the 
soft X-rays.

\section{METHOD}\label{method}

\subsection{Selection of microlensing events}

We base our work on the public data of the OGLE \citep{udalski1992}, MACHO \citep{alcock1993} and MOA \citep{muraki1999} 
collaborations available on the World Wide Web. Following the results of \citeauthor{st2010}, we select only events 
with time-scale longer than 100 days. 
The OGLE (Optical Gravitational Lensing Experiment) data\footnote{\it http://ogle.astrouw.edu.pl/.} were collected 
with the Early Warning System \cite[EWS, see][]{udalski2003} from 1998 to 2009 and correspond to the OGLE-II and 
OGLE-III phases of the survey. The number of events detected in the first year (1998) was 41, of which none had a duration 
longer than 100 days while e.g. in 2008 the number of events detected was 654 with 38 events longer than 100 days. 
The total number of events from OGLE is 4117, of which 177 percent fall in our range of interest.
The MACHO (MAssive Compact Halo Objects) survey data\footnote{\it http://wwwmacho.mcmaster.ca/.} were collected from 1993 
to 1999 and are comprehensive of 528 bulge events reported by \cite{thomas2005} plus the Red Clump Giants events reported by 
\cite{popowski2005}, totaling 567 events. Of these, 69 are longer than 100 days.
The MOA (Microlensing Observations in Astrophysics) survey\footnote{\it http://www.phys.canterbury.ac.nz/moa/} started 
in 2000, and in 2006 in entered in its second phase. While the number of events detected in 2000 was barely 8, 
the current detection rate is similar to that of the OGLE-III survey, i.e. $\sim 500-600\,\rm yr^{-1}$. 
The number of events detected by MOA up to 2010 is 2622, of which 268 show time-scales longer than 100 days.

\subsection{X-ray sources catalogs}

We searched for X-ray counterparts of our selected sample of ML events in the version 1.2 of the 
second XMM-Newton Serendipitous Source Catalog \citep[2XMM hereafter,][]{watson2009} and in the version 1.1 
of the Chandra Source Catalog \citep[CSC hereafter,][]{evans2010}.
The most recent version of the XMM 
catalog\footnote{\it http://xmmssc-www.star.le.ac.uk/Catalogue/xcat$\_$public$\_$2XMM.html} has been released in April 2010. 
It contains data of 191870 unique sources with a median flux (in the 0.2 - 12 keV band) 
$\sim 2.5 \times 10^{-14}\,{\rm erg\,s^{-1}\,cm^{-2}}$, with 20 percent of the sources having fluxes below 
$10^{-14}\,{\rm erg\,s^{-1}\,cm^{-2}}$. 
The CSC catalog\footnote{\it http://cxc.harvard.edu/csc/}, released in August 2010, covers approximately 320 square degrees 
of the sky at the $10^{-13}\,{\rm erg\,s^{-1}\,cm^{-2}}$ flux limit (0.5 - 7.0 keV band). The sky coverage drops to $\sim 6$ 
square degrees for a flux limit of $10^{-15}\,{\rm erg\,s^{-1}\,cm^{-2}}$ \citep{evans2010}. 
The total number of unique sources contained in the catalog is 106586.

\begin{figure}[b]
\centering
\includegraphics[width=0.6\textwidth]{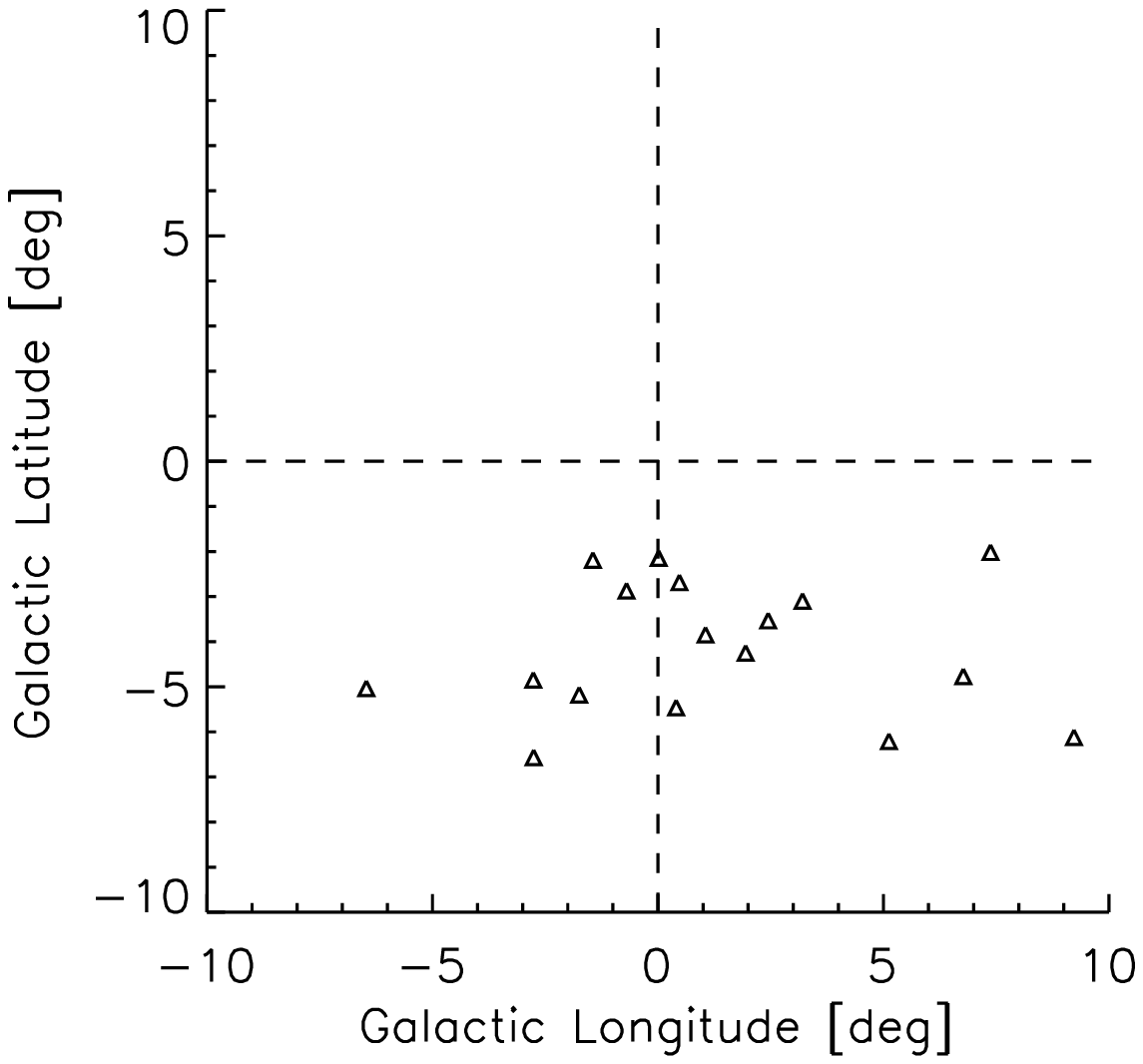}
\includegraphics[width=0.6\textwidth]{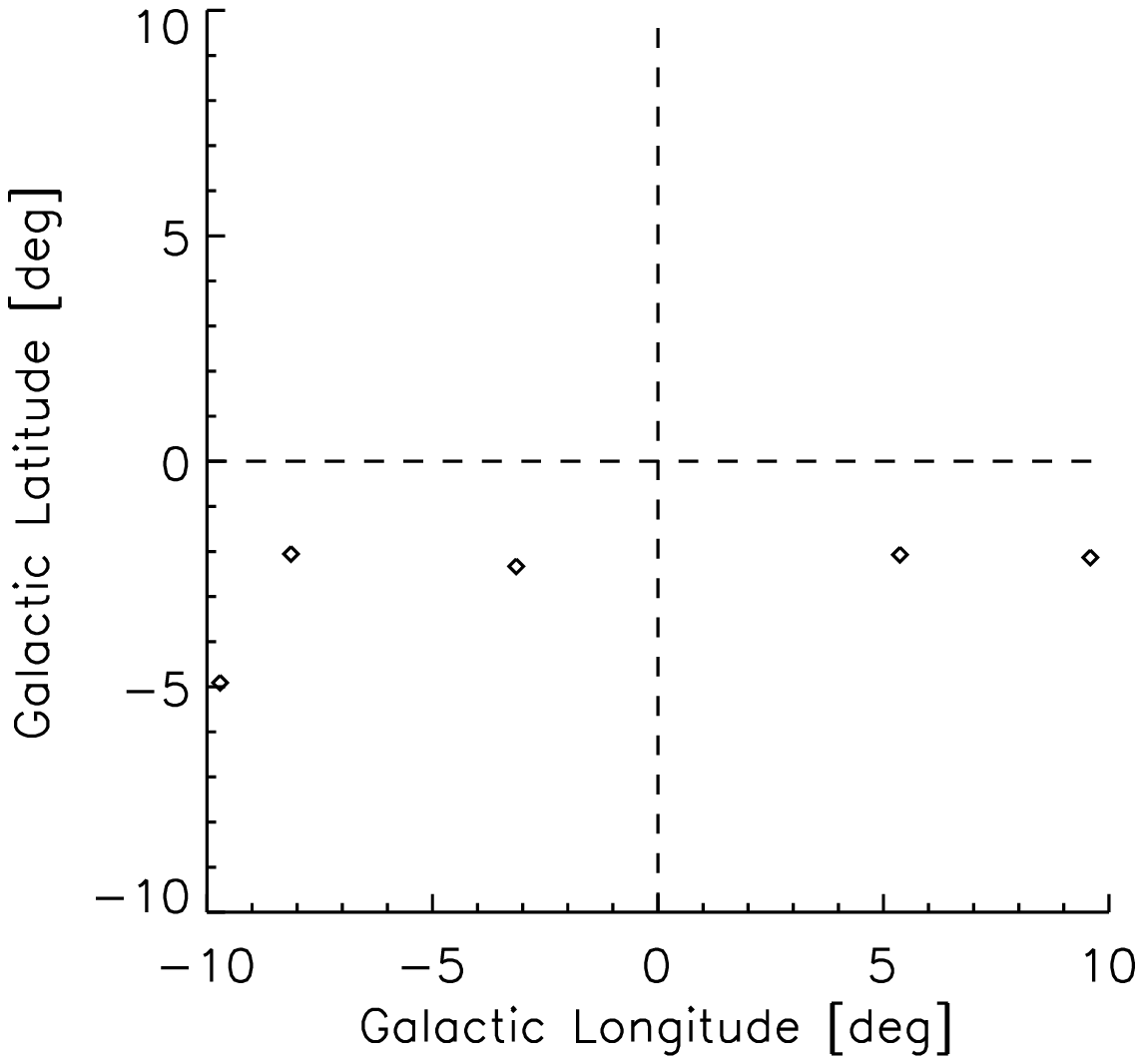}
\caption{Centroids of the pointings, in Galactic coordinates $(l,b)$, made by the XMM-Newton (top) and Chandra (bottom) 
satellites in the region of the sky monitored by the OGLE-III survey (see text).}
\label{fig-sky-xray}
\end{figure}

The OGLE-III survey monitored a region of the sky comprised between $l=-2^\circ$ and $l=-7^\circ$ and $b=\pm 10^\circ$, 
i.e. $\sim 100$ square degrees. The area covered by the MACHO and MOA-II surveys is roughly half of that of OGLE-III. 
\citep[e.g.][]{thomas2005}. The XMM-Newton and Chandra surveys made respectively 17 and 5 pointings towards the same 
area\footnote{multiple observations of the same region are considered as a single pointing} (see Fig. \ref{fig-sky-xray}) which, 
considering the $30'\times30'$ field of view of the EPIC cameras on board XMM-Newton and the $16'\times16'$ field of view of 
the ACIS cameras on board Chandra, correspond to $\sim 14$ and $\sim 1$ square degrees, respectively.

\subsection{Cross-correlation analysis}

Given the large uncertainties affecting the expected observational appearance of old isolated NSs and BHs, 
we do not apply any particular spectral or variability criteria on possible X-ray counterparts. 
Nevertheless, we get rid of spurious sources in the 2XMM catalog by selecting only those with $SUM\_FLAG = 0$, 
i.e. those for which none of the warning flags of the EPIC instruments were set. These flags indicate the probability of 
a detected source to be spurious.
For the CSC catalog there is no analogue flag, but the estimated number of spurious sources is expected to be 
less than one in every field with 100 kiloseconds of integration.
Thus, our matching criterion is based only on the positional coincidence of the long time-scale ML events with 
the X-ray sources of the 2XMM and CSC catalogs.

Our cross-correlation software computes the projected distance between the selected ML events and the entries 
of both the 2XMM and CSC catalogs. A positive match is found when a ML event lies with the $3\sigma$ error circle 
of an X-ray source.
Positional errors of single sources are taken from the respective catalogs. 
The positional error of 2XMM sources ($POSERR$ column) already accounts for systematic errors. 
The total $1\sigma$ uncertainty  is calculated as the squared root of the sum 
of statistical and systematic errors \citep{watson2009}

\begin{equation}
POSERR = \sqrt{RADEC\_ERR^2 + SYSERR^2}\,.
\end{equation}\bigskip

For CSC sources, to take into account the systematic error (0.16 arc-seconds) we use the equation suggested by the CSC team


\begin{eqnarray}
err\_ellip\_r_{0,tot} & = & 2.4477467 \times \\ 
  & & \times \sqrt{0.1669041 \times (err\_ellip\_r_{0,cat})^2 + 0.0256}\,. \nonumber
\end{eqnarray}\bigskip

Typically, the statistical error for on-axis CSC sources is $\sim\,0.2$ arc-seconds, while for off-axis sources at 
14 arc-minutes the statistical error is $\sim\,3.5$ arc-seconds.
We assume that positional errors of ML events are of $\sigma_{ML}\,\sim 1.5$ arc-seconds for all the events. 
Thus, the resulting radius of the error circle is assumed to be the root mean square of the X-ray source 
and ML event positional uncertainties

\begin{equation}
\sigma = \sqrt{\sigma_{ML}^2 + \sigma_{X}^2}\,.
\end{equation}\bigskip

\section{RESULTS}\label{results}

\begin{figure}[b]
\centering
\includegraphics[width=0.5\textwidth]{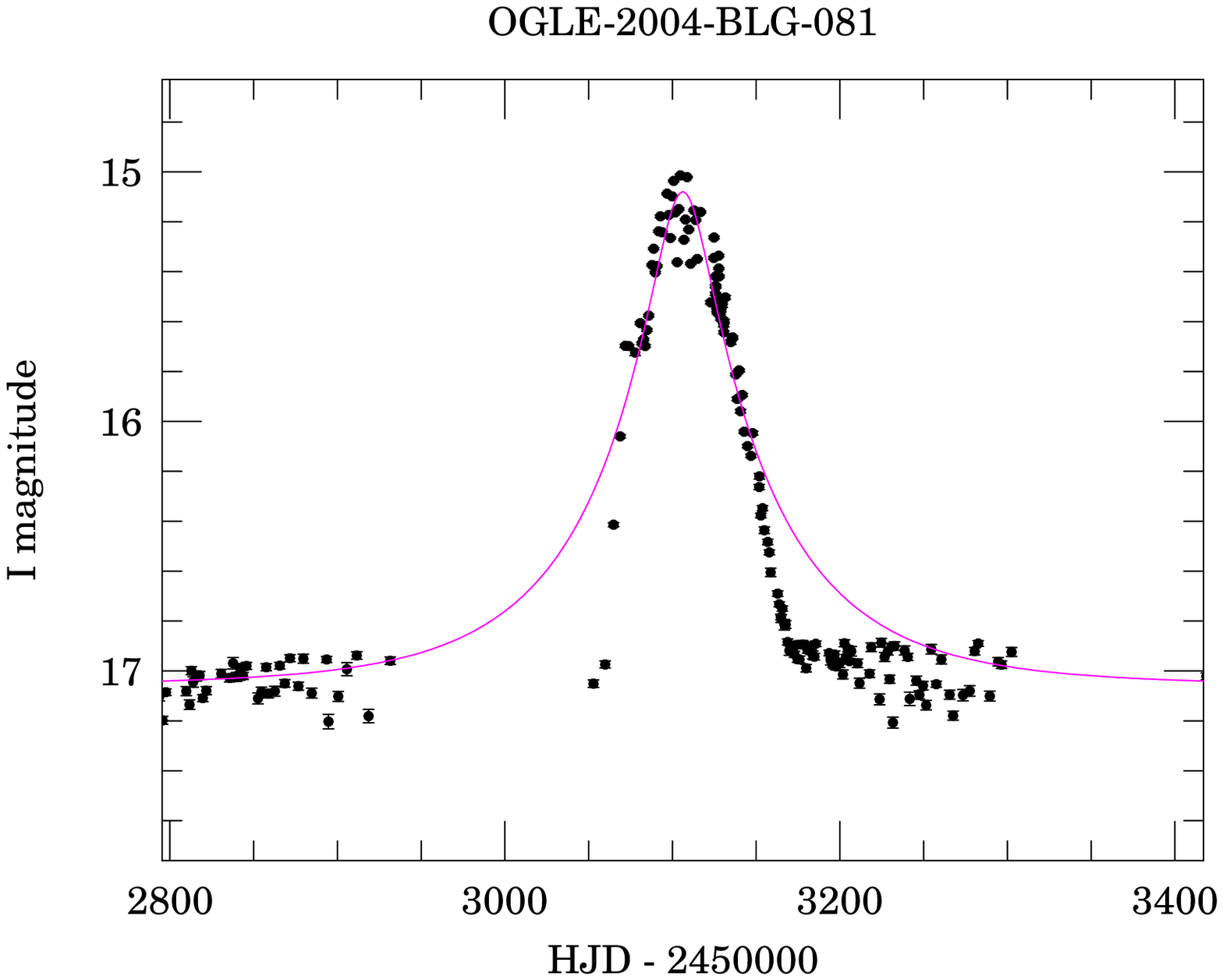}
\caption{Light curve of the event OGLE 2004-BLG-81. Solid line represents the best fit to the light curve 
assuming a standard microlensing model. Source: OGLE website.}
\label{fig-lcurve}
\end{figure}

The cross-correlation analysis returned a single positive match in the 2XMM catalog. 
The associated lensing event was observed in 2004 by both the OGLE and MOA surveys and is identified 
as OGLE 2004-BLG-81 and MOA 2004-BLG-3, respectively. The duration of the event reported by the OGLE team is $\sim 103.63$ days. 
However, the light curve is poorly fitted by standard lensing models (Fig. \ref{fig-lcurve}). \cite{wy2006} found that the baseline 
of the source star, i.e. the magnitude outside the ML event (I $\sim\,17$), has a suspected periodicity of $\sim 4$ days, 
thus pointing to an eclipsing binary. In particular, the shape of the folded light curve points to a contact binary system. 

To add more confusion, the MOA team reports a baseline $I\,\sim\,8$ (sic!), a duration of $\sim 6.73$ days and amplification 
very close to unity, $A\,\sim\,1.002$. However, a visual inspection of the stellar field does not confirm the presence 
of such a bright star, whose image would have been affected by diffraction. Thus we rely exclusively on the OGLE data to 
characterize the event.

The X-ray source associated to the ML event, 2XMM J180540.5-273427 (J1805 hereafter), has been serendipitously observed 
during a pointing of MACHO-96-BLG-5, another BH candidate detected through microlensing \citep{bennett2002, nucita2006}. 
The X-ray properties of the source have been retrieved with the XCat-DB web interface \citep{motch2009}. 
The total number of counts is $312.744 \pm 0.001$ (0.2 - 12 keV band) corresponding to a flux of 
$(3.39\, \pm 0.78)\, \times 10^{-14}\, \rm erg\,s^{-1}\,cm^{-2}$, which implies a luminosity, neglecting the photoelectric 
absorption of the ISM, of $\sim 3\, \times 10^{30}\, (d / 1\, \rm kpc)^2\, \rm erg\,s^{-1}$. If J1085 is the responsible for the 
magnification of a bulge star $(d \sim 8\, \rm kpc)$, then it should be placed at an intermediate distance and its 
X-ray luminosity should be lower than $\sim 10^{32}\, \rm erg\,s^{-1}$.

The positional uncertainty of the source is $\sim 2$ arc-seconds and it lies at $\sim 0.5$ arc-seconds from the position 
of the ML event. We report the fluxes on the different EPIC bands and the relative hardness ratios as given in the XCat 
database in Table \ref{tab-xray}.

\begin{table*}
\centering
\caption{Observed properties of the X-ray source 2XMM J180540.5-273427. The data are retrieved from the XCat database. 
Hardness ratios are calculated as described in \citeauthor{watson2009}, with the lower energy band corresponding to the position 
in the table.}
\begin{tabular}{c c c c c c}
\hline
\\
Energy Band & 0.2-0.5 & 0.5-1.0 & 1.0-2.0 & 2.0-4.5 & 4.5-12 \\
$\rm [keV]$ & & & & & \\
\\
\hline
\\
Flux & $0.002 \pm 0.001$ & $0.034 \pm 0.027$ & $0.298 \pm 0.050$ & $1.20 \pm 0.146$ & $1.73 \pm 0.762$ \\
$[{\times 10^{-14} \rm erg\,s^{-1}\,cm^{-2}}]$ & & & & & \\
\\
\hline
\\
Hardness ratio & $0.982 \pm 0.268$ & $0.738 \pm 0.141$ & $0.132 \pm 0.100$ & $-0.356 \pm 0.157$ \\
\\
\hline
\end{tabular}
\label{tab-xray}
\end{table*}

\section{DISCUSSION AND CONCLUSIONS}\label{discussion}

The small angular separation ($\sim 0.5$ arc-seconds) between the position of the ML event OGLE 2004-BLG-81 and J1805 
is well below the $1\sigma$ positional uncertainty of the X-ray source and makes the association highly likely. 
Thus, if J1805 is actually a BH, it would prove that ML surveys can detect isolated compact objects. 
However, there are a number of uncertainties that need to be addressed in order to not discard the claim.

First, the nature of the event reported by the OGLE and MOA surveys is unclear. As already pointed out, the shape of the 
light curve would rule out a ML event, even accounting for secondary effects like parallax or blending.
The fact that the source star is possibly a contact binary would indicate a cataclysmic variable (CV), i.e. an 
accreting white dwarf. This would imply that the event was in reality an outburst episode rather than a genuine gravitational lens.
An accreting white dwarf would easily explain the detected X-ray radioation as coming from the matter accreted by 
the degenerate star. However, the duration of the putative outburst and its lightcurve are unusual for this kind of sources 
\citep[see e.g.][]{butler1991,kuulkers2003}, thus challenging the CV hypothesis.

In alternative, it has been suggested that the lensed star is a chromospherically active variable \citep{bernhard2009}, 
possibly of the RS Canum Venaticorum  (RS CVn) type. This class of variables is known to show periodic variations which are 
thought to be related to the active regions on the surface of the star or ellipticity of the star itself. 
These effects can mimic the lightcurve of an eclipsing binary and explain the observed modulation of the baseline. 
Furthermore, RS CVn stars are known X-ray emitters, with luminosities of $\sim 10^{31} \rm erg\,s^{-1}$, 
and they also show flaring activity at both optical and X-ray wavelengths.
Yet, also in this case the amplitude, shape and duration of the optical event are unusual for RS CVn stars 
(S.N. Shore, private communication).

Aside from the possible classification of the optical event, the nature of the X-ray source J1805 is poorly constrained 
as well. The small number of photons collected does not allow a good characterization of the spectrum. 
The hardness ratios are positive at low energies and change sign between 2 and 4.5 keV. 
This fact would suggest a hard or highly absorbed spectrum. A hard spectrum would rule out a NS as accreting object since its 
temperature is expected to be below $\sim 1\, \rm keV$. Thus, if J1805 is an isolated compact object, it should be a BH.

In conclusion, in this paper we performed a cross-correlation analysis of long time-scale microlensing events 
with the soft X-ray sources of the recently released XMM-Newton and Chandra catalogs. We report a positive match between 
the ML event OGLE 2004-BLG-81 and the X-ray source 2XMM J180540.5-273427. 
The nature of both the ML event and the the X-ray source are still unclear. Follow-up optical and X-ray observations 
of the field of 2XMM J180540.5-273427 at high angular resolution could help to characterize the source star and 
the X-ray source as well. 

We stress that, even if the association between OGLE 2004-BLG-81 and 2XMM J180540.5-273427 would not be confirmed, 
the argument that a large fraction of long duration events are related to BHs is still valid.
 As shown in Section \ref{spectra}, and in particular in Equation \ref{eqn-lx}, 
the expected flux from an isolated BH or NS depends on many unknown quantities:
the velocity of the collapsed object, the density of the ISM, the photoelectric absorption etc. 
Therefore, it is not straightforward to use the absence, or paucity, of correlation between
X-ray sources and ML events as a constraint on so many parameters.

In the next years we expect a substantial enlargement of the catalogs of ML events and X-ray sources. 
In particular the eROSITA mission, which should be launched shortly, will make a survey of the entire sky, 
which in the soft X-ray band (0.5 - 2 keV) will be 30 times more sensitive than ROSAT. 
At the same time, deep systematic exposures with Chandra and XMM-Newton, of the most promising ML events, chosen 
mainly on the basis of the duration, will also help to set stronger constraints on the flux of accreting isolated 
compact objects.


\begin{acknowledgements}
The work presented in this paper was made possible thanks to the public data from the OGLE, MACHO and MOA microlensing surveys
and from the X-ray source catalogs of the XMM-Newton and Chandra satellites. 
We thank the anonymous referee for helpful comments and suggestions which improved the previous version of this paper. 
NS also thanks A. Paizis for useful discussion. The work of NS is supported by ASI-INAF through grant I/009/10/0.
\end{acknowledgements}




\end{document}